\documentclass[12pt,preprint]{aastex}
\usepackage{amsmath}
\usepackage{amssymb}
\usepackage{amsthm}

\shorttitle{Nature of the blazar CGRaBS\ J0211+1051}
\shortauthors{Chandra et al.}

\begin{document}
\title{Understanding the nature of the blazar CGRaBS~J0211$+$1051}
\author{Sunil Chandra, Kiran S. Baliyan and S. Ganesh}
\affil{Physical Research Laboratory, Ahmedabad,  India 380009}
\email{sunilc@prl.res.in}
\and
\author{L. Foschini}
\affil{INAF, Osservatorio Astronomico di Brera, Via E. Bianchi 46, 23807, Merate (LC), Italy }

\begin{abstract}
The source CGRaBS~J0211$+$1051 (MG1 J021114+1051, z=0.20) flared up on 2011 January 23  in high-energy $\gamma$-rays as reported by {\it Fermi}/LAT. This event was followed by the increased activity at the UV, optical and radio frequencies as detected by  the observing facilities worldwide. The source also showed a high and variable optical polarization based on which it was proposed to be a low-energy peaked BL Lac Object (LBL). Present work reports first comprehensive  multi-wavelength study of this source using data in the radio, optical, UV, X- and $\gamma$-rays and optical polarization. Using these multi-wavelength data on the source,  we have estimated various parameters and verified it's classification vis-a-vis blazar sequence. Multi-waveband light-curves are used to discuss flaring events of 2011 January  in an attempt to address the nature of the source and pinpoint the possible physical processes responsible for the emission. The light-curves show variations in the high energy $\gamma$-rays to be correlated with X-ray, UV and optical variations, perhaps indicating to their co-spatial origin.   Our optical  data, quasi-simultaneous with UV (Swift-UVOT) and X-ray (Swift-XRT) data, enabled us to trace low energy (Synchrotron) component of the spectral energy distribution (SED) for  CGRaBS J0211+1051, for the first time. The  SED shows the synchrotron peak to lie at $\sim 1.35\times10^{14}$~Hz, confirming CGRaBS~J0211+1051 to be an LBL. Some other parameters, such as local magnetic field ($\sim 5.93 $ Gauss) and black hole mass ($\sim 2.4 \times 10^{8} M_{\odot}$) are also estimated which are in agreement with their typical values for  the blazars.  Based on the present study, identification of the Fermi/LAT source,  2FGL J0211.2+1050,   with its BL Lac counterpart CGRaBS J0211+1051 is confirmed.
      
\end{abstract}


\section{Introduction}

Blazars are the extreme class of active galactic nuclei (AGNs). They show variable continuum emission over the whole energy spectrum along with  high ($\geq 3 \%$) and variable polarization in the radio and the optical energy bands. According to the AGN unification model, blazars have a relativistic jet of plasma,  perpendicular to the accretion disk,  emanating from very close to the central black hole and aligned  at a very small angle  to the line of sight($\lesssim 10^{\circ}$) \citep{UrryPadovani1995, UrryPadovani2000}.  Due to the relativistic aberration and boosting effects, the emission from blazars is dominated by the jet. The  variability in  their emitted flux provides  an important tool to study the central engine in AGNs since these remain spatially unresolved even by the best observing facilities of the day. The observed timescales of the flux variability provide  clues to the size of the emission region and  the processes responsible for the emission.
         
 The spectral energy distribution (SED) of the blazars has a characteristic double-peaked shape with a low frequency component peaking somewhere in sub-mm to X-rays energy band, and a high frequency component, peaking at the MeV-TeV energies. The low energy component of the SED is well explained as the synchrotron emission from relativistic electrons in the jet \citep{UrryMushotzsky1982}, while the physics behind the high energy component is not yet  well understood. It is proposed that the high energy flux is produced by the inverse Compton scattering of the  low energy seed photons by the highly energetic particles (electrons/postitrons). The origin of these seed photons can either be the synchrotron emission itself (Synchrotron Self Compton, SSC; \citet{GG85, BloomMarscher1996, SokolovMarscher2004}) or the sources external to the emission region, e.g., accretion disk, broad line region, torus etc (External Compton, EC). Several models have already been proposed to explain source of the seed photons responsible for  the high energy emission in blazars  \citep{Dermer1992, Sikora1994, Blazejowski2000, sikora2009, Agudo2011}. In order to constrain the models for the high energy flux generation in blazars, study of  the light-curves and the SEDs of a sample of blazars is needed using  long term, simultaneous multi-wavelength data. 

   Depending upon the position of the synchrotron peak in SEDs, blazars are sub-classified into a sequence; flat spectrum radio quasars (FSRQ), radio-selected BL Lac objects (or LBL), and X-ray-selected BL Lac objects (or HBL) \citep[and references therein]{UrryPadovani1995, Fan1997, HeidtNilsson2011}. These sub-classes are known to have some intrinsically different properties. For example, their bolometric luminosity decreases from FSRQ to HBL as does the dominance of the  $\gamma$-ray emission \citep{Fossati1998, Sambruna2009}. Similarly, LBLs are reported to have, on the average, higher degree of polarization (DP) and amplitude of variation than the HBLs \citep[e.g.] [and references therein]{Andruchow2005, Tommasi2001, Fan1997, Jannuzi1993}. \citet{Fan1997} ascribe this difference in the DP to the differences in their beaming  with LBLs showing stronger beaming. Recently, \citet{HeidtNilsson2011} found only a marginal difference in the polarization behaviour of the LBLs and the HBLs as inferred from their sample of probable blazar candidates taken from the Sloan Digital Sky Survey. However, their inference could be affected by the low statistics as they considered 8 LBLs and 37 HBLs in their sample. On the other hand, \citet{Andruchow2005} and \citet{Ikejiri2009} report, based on the studies of their samples, that LBLs generally have a higher DP than HBLs. 
           
       In an earlier work, \citet{Chandra2012} have discussed intra-night as well as inter-night variations in DP and position angle (PA) for the blazar CGRABsJ0211+1051.  This source was detected by EGRET on-board Compton Gamma Ray Observatory (CGRO) as an unidentified source, 3EG  J02115+1123 (RA= 34.00$^\circ$, DEC=11.38$^\circ$; \citet{Hartman1999}) with  1.08$^\circ$ error circle.  The first and second Fermi-LAT catalogs\citep{Abdo2010, nolan2012} have several sources in this error circle including the source 1FGL J0211.2+1049/2FGL J0211.2+1050  which was found associated with  source MG1 J021114+1051 \citep{Griffith1991} from First MIT-Green Bank 5 GHz Survey and 87GB 020832.6+103726 (1987 Green Bank Radio Survey, \citet{Gregory1991}). The source CGRaBS J0211+1051 was detected in Candidate Gamma-Ray Blazar Survey\citep{Healey2008} and  due to its featureless optical spectrum, was categorized to be a BL Lac object \citep{Healey2007, Lawrence1986, Snellen2002}. Swift\citep{Burrows2005} detected a single X-Ray source within 1.5' radius centered at CGRaBS J0211+1051. \citet{MeisnerRomani2010} reported a redshift of $0.2 \pm 0.05$ for CGRaBS J0211+1050 (MG1 J021114+1051, 2FGL  J0211.2+1050), consistent with the one reported for 3EG  J02115+1123.  Very recently, CGRaBS J0211+1051 has shown brightening in near infrared (H band= 11.45 mag \citet{Carrasco2013}) and high optical polarization ($\approx 22\pm 4$ \%  \citet{Grigoreva2013}).  During 2011 January 30-February 03 observations \citet{Chandra2012} found that the source showed significantly high and variable   ($9-21 \%$)  degree of polarization on the time-scales of hours. Based on that, authors proposed this source to be an  LBL or RBL. However a true test for such classification is only provided by  the location of the synchrotron peak in its SED.   

In the present study, we construct the SED of  CGRaBSJ0211+1051 using the multi-wavelength data obtained during 2011 January 24 to February 03 from various observatories  and discuss the nature of the source. This paper is organized as follows. The next section discusses the observations and analysis of the data. In section $3$ we discuss light-curves and SED of the source and put forth the conclusions drawn from this work in section $4$.

\section{Observations and Data Analysis}

 In order to study the behaviour of blazar {\it CGRaBSJ0211+1051}, we generated the light-curve and SED using quasi-simultaneous   data in all available energy bands. For SED, all the data used are almost simultaneous, excluding few points in radio and infrared region (WISE: http://wise2.ipac.caltech.edu/docs/release/allsky/expsup/sec16b.html). The sub-mm \citep{Planck11}  data are from 2012 observations \citep{Ade2013}.  The high energy $\gamma$- and X-ray data are taken from Large Area Telescope(LAT)  \citep{Atwood2009} on-board {\it Fermi} and XRT on-board {\it Swift} space based observatories, respectively. Recent version of ScienceTools (version v9r27) is used to analyze LAT data. The data from the X-ray telescope (XRT; \citet{Burrows2005}), and  optical/ultraviolet monitor (UVOT; \citet{Roming2005}) are processed and analyzed using HEASOFT version 6.12 with calibration database as updated on 2011 Aug 25. For constructing the SED,  the fluxes at radio frequencies (8.4 GHz, 4.85 GHz, 4.775 GHz and 1.4 GHz) are retrieved from NASA Extragalactic Database (NED) and  do not belong to the same epoch.  The 2cm flux from MOJAVE database \citep{ListerMOJAVE2009}, observed on 2011 February 27, is also used. In the following  we summarize  techniques used for the analysis of various data-sets.  

\subsection{Optical Polarization and R-band observations from MIRO}

The polarimetric observations for this source were carried out using PRL made optical photo-polarimeter (PRLPOL) as a backend instrument mounted at the f/13 Cassegrain focus of 1.2 m telescope of the  Mt. Abu Infra-Red Observatory (MIRO). The detailed information about the instrument used and measurements made during 2011 January 30-February 03 are given by \cite{Chandra2012}.   The  online data reduction provides DP,  error in DP, position angle (PA) of the polarization and other parameters,  as output.  The nightly averaged values of DP and PA along with their respective standard deviations as reported in Table 2 by \citet{Chandra2012} are plotted in Fig. 1(e) and (f), respectively. 

The  optical observations in the  R-band were performed using recently installed 0.5 m aperture optical telescope, known as Automated Telescope for Variability Studies (ATVS), at MIRO.  The field of view and plate scale of the system are $13'.5 \times 13'.5$ and 0.79 arcsec/pixel, respectively.  The  blazar CGRaBS J0211+1051 was observed during 2011 Jan 30-31 and 2011 Feb 01 \& 03.  The field of view of the CCD is large enough to accommodate a number of comparison stars in the field of the source for differential photometry. However, in the vicinity of source there are no known standard stars which can be used for calibration. We, therefore, performed aperture photometry on all the stars present in the field. The observed magnitudes of these stars are corrected using the photographic plate magnitudes from USNO catalog after appropriate filter conversion. We tested ten stars in the field with brightness comparable to the source, for their stability in flux during the course of monitoring. Three out of ten stars are found  appropriate to be used as standard comparison stars for the present analysis. We used one of the three, close in brightness to the source as comparison star and rest as control stars to  construct the differential light curves in order to check the source for  variability. The observed  magnitudes of CGRaBS J0211+1051 were corrected using the  averaged magnitudes of all the three stars. The extinction correction is performed as prescribed by \citet{Cardelli1989}. The conversion of magnitude to energy flux was performed using appropriate factor and zero point flux as described by \citet{bessel1979}. Fig. 1(d) shows the R-band flux as a function of  time. Table 3 presents the magnitudes derived for the source and other three comparison stars in the field.  
     
\subsection{Swift Observations}

We have made use of HEASARC archival database for this source during the course of 2011 January flaring period (MJD 55586-95). The observation Ids of the data used for the present analysis along with their respective exposure times are listed in Table 1.  {\it Swift} started following this source just after the report of an intense flaring activity in $\gamma$-rays by LAT onboard {\it Fermi}  \citep{AmmandoAtel2011} on January 25, 2011. The data from the instruments on-board {\it Swift}, namely, XRT and UVOT are downloaded from the website and analyzed for the present study. The latest version of HEASOFT package (v6.12) with a calibration database updated on August 25, 2011, is used for the analysis.  In the following we describe details of analysis adopted for the data from various detectors on-board {\it Swift}.
 
\subsubsection{XRT}  
     The level 2 cleaned event files are generated using standard procedures as recommended in the manual by instrument team. The default screening parameters are used. For the PC mode grades 0-12 and for WT mode grades 0-2 are selected by using Ftool {\it xselect}. The background light curve and spectrum are generated after using appropriate region filtering. In this case we have taken a circular area of 15 pixel radius around the target as the source region and  four source free regions in the neighborhood of the target, each of 45 pixel radius,  as background. The required ancillary response matrix is generated by using task {\it xrtmkarf} followed by {\it xrtcentroid} task. The response matrix file provided with the CALDB distribution is used for further analysis. The spectrum thus obtained  is then fitted to generate the lightcurve and SED. 
   
    The spectral fitting was done in the energy band between 0.2 to 10.0 keV using XSPEC (version 12.7.0) package distributed with HEASOFT 6.12. The simple power law along with the Galactic absorption gives the best fit for almost all the observations of interest. The model parameter, interstellar column density, $N_{\rm H}$, is kept fixed at a value of $5.5\times 10^{22}$~cm$^{-2}$,  \citep{Kalberla05}. Table 2 summarizes the values of various parameters obtained from the spectral fitting for different epochs. We do not see any significant variation in the photon index for different observations implying that the source remains in same spectral state during the course of monitoring. We, however,  notice variation in flux as indicated by the light-curve [Fig. 1b]. 

    For constructing SED, we estimated the photon flux in several small energy bins using interactive plotting utility (IPLOT), a part of PGPLOT, while fitting with XSPEC. The Galactic extinction correction is done in the following manner. First of all, modeled photon flux is calculated for all small energy bands using IPL tool with $nH$ parameter as described above. The same procedure is repeated with $N_{\rm H}=0$ which assumes no absorbing material in that particular line of sight. The ratio between absorbed and un-absorbed model photon fluxes gives the absorption factor which is then used to correct the observed photon flux calculated by IPL in order to get intrinsic photon flux. The galactic extinction corrected photon fluxes respective to different energy bins are then converted into energy fluxes with appropriate conversion factor before  using to construct SED (Fig.2).     

\subsubsection{UVOT}

      UVOT snapshots with all the six available filters, V (5468 $\AA$), B (4392 $\AA$), U (3465 $\AA$), UVW1 (2600 $\AA$), UVM2(2246 $\AA$) and UVW2 (1928 $\AA$) for all the obsIds (Tables.1), were integrated with the {\it uvotimsum} task and analysed  using  {\it uvotsource} task, with a source region of 5$\arcsec$, while the background was extracted from an annular region centered on the blazar with external and internal radii of $40 \arcsec$ and 7", respectively \citep{FoschiniL10}.  The observed magnitudes from all obsId are then corrected for extinction according to the model described in \citet{Cardelli1989}. The magnitudes thus obtained are converted to energy flux (erg cm$^{-2}$ $s^{-1}$ \AA$^{-1}$) using the following equation:
  \begin{equation}
F_{\lambda} = FCF \times 10^{(ZPT-m)/2.5}
\end{equation}
where {\it ZPT} is zero point flux, {\it FCF} is the flux conversion factor (erg$^{2}$~cm$^{-2}$~count$^{-1}$~\AA$^{-1}$) and {\it m} is the observed magnitude in a particular filter. These standard values are taken from instrument calibration database (CALDB, \citet{Poole2008}). The light-curves are then constructed using the flux values for the filters V, B, U, UVW1, UVM2, and UVW2 for different ObsIds (Fig. 1c). The UVOT flux values averaged over whole observing period, are used to construct the SED (Fig.2).

\subsection{Fermi/LAT Observations}
The energy coverage of LAT on-board {\it Fermi} Gamma-ray observatory is broad enough ($\approx$ 20 MeV to $>$ 300 GeV) to cover the part of blazar SED which is assumed to be mainly contributed by the inverse Compton processes in the jet. CGRaBS~J0211$+$1051  was first reported in outburst state at 0.1-100 GeV by \citet{AmmandoAtel2011}. In order to investigate the high energy emission before and after the flaring period along with the outburst state, we analyzed the data ranging from MJD 55548 to MJD 55610 (approximately two months). Only a part of this, coinciding with the swfit observations (MJD 55586-95), is shown in Fig. 1. 
 The PASS7 photon data with region of interest (ROI) of $15^{\circ}$ are analyzed using latest version of ScienceTool (v9r27) and instrument response functions (IRFs),  P7SOURCE\_V6. All the events with zenith angle $\ge 100^{\circ}$ are discarded to avoid the contamination due to the $\gamma$-ray bright earth along with the appropriate gtmktime filters.  The unbinned likelihood analysis has been used to construct the source energy spectrum. For this,  first of all,  coordinate, time, energy and region selections are performed on the raw event file to avoid unwanted contributions. The output file of previous step is again corrected for detector live time.  A source model is constructed using the contributory python script {\it make2FGLxml} incorporating the latest {\it Fermi}-LAT catalog {\it gll\_psc\_v07.fit}, diffuse background components {\it gal\_2yearp7v6\_v0.fits} and extragalactic background {\it iso\_p7v6source.txt}. The Galactic diffuse emission model is generated using the GALPROP package, available online as a contributory file, while the extragalactic one is described by a simple power law \citep{Abdo2009}.

         We have adopted the methodology  of \citet{FoschiniL10, FoschiniL11}, briefly discussed here. First, the unbinned likelihood analysis was performed on the complete data, spaning over energy band 0.1 to 100 GeV in order to determine the best source model. For this event the {\it powerlaw2} model is found to be the best fit source model as inferred by high test statistics (530.9). The power law index value for this fitting is  $2.03 \pm 0.06$. The source model with the power law index frozen to that value in the fit is used for constructing the light curve with 3-day binning. The flux values with TS $\textless 9 $, equivalent to $\sigma \textless 3$, are discarded from the final lightcurves.

       The data corresponding to the time interval MJD 55580 to MJD 55596 are  used to construct the blazar SED. For extracting SED,  the event file is binned in several energy segments (e.g., 100MeV - 500MeV,  500MeV - 1GeV, 1GeV - 5GeV, 5GeV - 50GeV, 50GeV - 100 GeV and 100GeV - 200GeV) and likelihood analysis is performed over each energy bin, individually, to get energy flux for respective energy bands. The source model used for this part is similar to that used for complete energy band except the energy interval. For each energy bin, the source under investigation and all nearby sources in the region of interest  are described by one parameter representing the integral flux in that energy bin. The diffuse background components are modeled with one single parameter describing the normalization. The upper-limit estimation is done for last two energy bins as TS is always less than 9 for these bins. 

The following tools provided as a part of the software distribution are used for the analysis done here. The {\it gtselect} and {\it gtmktime} are used for event selection and live time correction, respectively. The {\it gtltcube}, {\it gtdiffrsp} and {\it gtexpmap} are used for generating livetime cube , Galactic diffuse response, and exposure map respectively. The likelihood analysis is performed using tool {\it gtlike}. It provides the test statistics for source model fit along with the other model parameters. 
  
\section{Results and Discussion}
The blazar CGRaBSJ0211+1051 is an interesting source which has been brightening since 2005 \citep{Chandra2012} from the levels of 15.5 mag in V band \citep{Djorgovski2011} and had undergone a strong flare during 2011 January 25 - February 03. It was reported to show bright state in almost all wave-bands while {\it Fermi} $\gamma$-ray photon flux was reported to rise by 25 times the yearly averaged values \citep{AmmandoAtel2011}. We used the available data on the source along with our own observations to produce light curves and spectral energy distribution to understand the nature of the source and to ascertain its actual classification. In the following we discuss the results obtained from this study and available information from the literature.

\subsection{Multi-wavelength light-curves and optical polarization }

 Figure. 1 shows the multi-wavelength light curves for the blazar CGRABsJ0211+1051. The panels (a) and (b) show the $\gamma$-ray and X-ray flux variations, respectively. The panel (c) contains UV/optical light-curves  as obtained from UVOT on-board {\it Swift}. The panel (d) shows the R-band light-curve obtained from the  EMCCD, mounted at ATVS, data. The last two panels, (e) and (f), show the variations in nightly averaged degree of polarization and position angle as discussed in \citet{Chandra2012}.  In the following we discuss each component individually and in relation with source behaviour in other wavebands.

 The visual inspection of Fig. 1(f)   clearly shows that emission was highly polarized $(DP \sim 21.05 \%)$ on MJD 55591.38 (2011 January 30),  indicating a highly aligned magnetic field in the emission region. The DP gradually decreased to  $10.63 \%$ on 2011 February 01 (MJD 55593.44) at a rate of $5 \%$ per day.   DP then increases again at a relatively slower rate ($2.42 \%$ per day) over next two days. The PA follows the similar trend with a difference that it decreases on 2011 February 03. The variation in DP can be explained using  intrinsic models such as shock in jet model, fresh injection of matter in the jet etc., depending on the time-scale and nature of changes observed. The change in PA is unlikely to be caused in a straight, uniform axially symmetric, matter dominated jet just by shock compression of plasma in the emission region. The interaction with a perpendicular shock moving along blazar jet can align the tangled magnetic field in the emission region, hence enhancing DP but might not result in any significant change in PA \citep{Abdo2010}. Such a change in the PA (on day 1,  3 and 5) indicates either the fresh injection of material in the jet or a change in the jet geometry. The flipping nature of  PA around MJD 55594.44 appears  interesting because of opposite behaviour in DP.  Such trend has been noticed in other blazars when DP is seen rising accompanied with sharp drops in PA.
  
           The R band light curve (Fig.1d) shows that the source has seen its brightest moments before 2011 January 30  with the flux decreasing during 2011 January 30 - February 03,  in agreement with the trend reported by  \citet{NesciAtel2011} ($R_c \sim 13.37$ on 2011 January 27). The R band light-curve also indicates to a small rise in flux between MJD 55592.4 and MJD 55595.4 around which DP and PA are showing opposite behaviour. This might be indicative of a  small flaring activity, enhancing the R-band flux and DP  followed by a drastic change in PA. Since timescale is very small ($\sim$ 2 days),  this flare can not be due to global bending of the jet. It  possibly favors a fresh  injection of plasma in the emission zone as cause for the change.
 The UV-Optcal light-curves (Fig. 1c)  show  similar behaviour in all bands, albeit with  a weak colour dependence.  A mild bluer when brighter and redder when fainter behaviour seen is consistent with the shock-in-jet model. We notice about 20\% increase in  U-band energy flux  during MJD 55586.8 to MJD 55590.4 (2011 January 26 to 29).  In all optical and UV bands, the source had maxima and minima around MJD 55590 and MJD 55595.5 with values in U, B, V bands as 9.58, 11.2, 14.3 mJy and 7.02, 8.1, 10.4 mJy, respectively. R-band flux peaked on MJD 55591.4 (16.67mJy)  with a minimum on MJD 55595.4 (14.14mJy). The  delay of about one day in the maxima and minima  as seen in our R-band vis-a-vis UBV bands is  possibly due only to sampling time.
As shown in Fig 1(b), XRT-Swift X-ray energy flux goes up by about 50\%   (from 0.18 to 0.27 mJy) within about 4 days (MJD 55586.4 to 55590.4) and  then drops by 0.04 mJy within next 2.8 days.  The X-Ray (0.2-10.0 keV) light-curve  follows the similar trend as seen in UV bands apart from an increase in X-ray flux towards the end while fluxes in R and UV bands decrease.  The maximum (0.28 mJy) and minimum (0.16mJy) in the X-ray light curve occur  on MJD 55590.08 and MJD 55587.02, respectively.

On 2011 Jan 25 \citet{AmmandoAtel2011}  reported the source to have highest ever $\gamma$-Ray flux on 2011 January 23. The $\gamma$-Ray lightcurves shown in Fig 1(a) are obtained from Fermi/LAT observations with an inclusion of flux corresponding to   TS $\ge$ 9 ($\sigma \sim 3$) and 3-day binning.  A significant rise around 2011 January 23  and again on 2011 January 29 was  noticed.   The   January 29 peak flux  decays slowly with time reaching a minimum on MJD 55595.  This flare appears to be different from the one around 2011 January 23, decaying part of which might have overlapped with rising part of the second (January 29) flare but the peak photon flux (4.3$\times10^{-7}photons/cm^2/sec$) is at almost same level.
 
  Fig 1 shows that  $\gamma$-ray, X-ray, $UV$ and optical fluxes vary largely in unison during 2011 January 25 to February 02 with light-curves in all the bands peaking somewhere near 2011, January 29.
 Though there are no polarimetric observations on 2011, January 29,  a polarization value of 12\%  on January 28 was reported by \citet{Gorbovskoy11}, followed by our measurements of $\sim 21\%$ DP  two days later on January 30. It clearly shows  a trend  of rapid increase in DP during January 28 - 30 followed by equally rapid drop with January 31 recording a value of 12.8\% indicating DP to also peak sometime on  January 29.  While PA follows DP during January 30-Feb 2, we do not know how it behaved near the peak  (January 29) as there are no measurements on January 28 for PA. As mentioned earlier, from 2011 February 01 (MJD 55593.4) DP  increases slowly while PA shows an increase upto February 02 and then decreases.  Interestingly, flux in the R-band also shows a mild rise on February 02 and then a sharp fall by about 10\%,  just like PA.  A slight increase in X-ray light curve  is also noticeable at this epoch. The one day averaged {\it Fermi}/LAT flux also shows an enhancement on February 02 (MJD 55594). It might be indicative of a flicker at around MJD 55594.4 caused by the inhomogeneity in the jet, leading to enhancement in the flux and polarization accompanied with change in PA. However,  nothing can be said of  UVOT fluxes due to the lack of the  UVOT pointed observations at this epoch. 

 We, therefore, conclude that in totality, variations in fluxes in all the bands appear to be simultaneous in nature, indicating co-spatiality of the emission at all  wavebands considered here.  Another important consequence of this result is that the Fermi source 2FGL J0211.1+1050, as well as x-ray counterpart are identified with BL Lac object CGRaBS J0211+1051.  However, one can not miss the differences in the nature of short term small scale fluctuations in the light curves for different energies, implying the presence of small scale inhomogeneities in the physical conditions across the source.

\subsection{Spectral Energy Distribution (SED)}
 In order to construct  SED for CGRaBS\ J0211+1051, we used data discussed above alongwith MOJAVE (2 cm) and other (8.4 GHz, 4.85 GHz, 4.775 GHz and 1.4 GHz) radio band data  from NED website. We analyzed the data  for the duration MJD 55580-MJD 55596 (2011 January 20 - February 04) which are quasi-simultaneous in nature. We combined the fluxes in respective energy bands for the duration data was available for a particular band, after correcting for the Galactic extinction and other aberrations. We, therefore, combined 15 days data for various UVOT/X-ray and $\gamma$-ray energy bands from Swift and LAT  while  R-band optical data was combined for 5-days (2011 January 30- February 03). The  SED is plotted  as $\nu F\nu \  v/s \ \nu$ on logarithmic scale in Fig 2. The filled circles in the first peak represent  the fluxes in various Radio bands while filled diamonds, triangles and squares represent sub-mm\citep{Planck11}, infrared (wise) and optical UV fluxes. The  X-ray 0.3-10 keV fluxes from XRT and high energy LAT/Fermi $\gamma$-ray fluxes are shown by open squares and inverted triangles.  The last two data points (open circles)  are upper limits on the $\gamma$-ray fluxes corresponding to TS values less than 9. It is interesting to see from Fig.2 that CGRaBS J0211+1051 has a strong IC component. A color version of this figure is available online.

 From the nature of the spectra, it is clear that the synchrotron peak falls some where in optical/near IR region. To have  a better estimate of the low energy (synchrotron) peak frequency for CGRaBS J0211+1051, 
 following parabola was fitted to the lower energy component of SED.

\begin{equation}
\log(\nu L_{\nu}) = A \, [\log(\nu)]^2 + B \,  \log(\nu) + C
 \end{equation}

where A, B and C are constants  estimated using general non-linear model fitting algorithm freely available in statistical software $R$.  The best fit values of parameters A,  B and C are,  -0.23 $\pm$ 0.04, 6.17 $\pm$ 1.03 and -52.48 $\pm$ 6.01, respectively. The corresponding synchrotron peak energy flux and  the peak position  are $\sim$ 5.21 $\times$ $10^{-11} erg/cm^{2}/s $ and $\sim$ 1.35 $\times 10^{14}$ Hz, respectively.  It confirms the position of the peak of low energy component (synchrotron) of this BL Lac  in near infrared region, categorizing the source as ''low energy peaked (LBL)" (or RBL), supporting  the suggestion made by \citet{Chandra2012} based on their polarization measurements, that this source belongs to the class of  low energy peak blazar (LBL).  The present study, therefore, gives credence to the idea that blazars can largely be classified based on their polarization properties.

     Now, using the quantities estimated above, we can further quantify some of the other parameters for this source.  Using the values of the low energy peak flux and synchrotron peak frequency in the following expression by \citet{bottcher2007}, one can estimate the co-moving magnetic field in the emission region. 
 
  \begin{equation}
B_{eB} = 9D_1^{-1} \left [ \frac{d_{27} ^{4} f_{-10} ^{2} e_{B} ^{2}} {(1+z)^{4} \epsilon _{sy,-6} R_{15} ^{6} (p-2)} \right ]^{1/7} G
\end{equation}
where
synchrotron peak flux;
\begin{displaymath}
 f_{-10} = \frac{ f_{\epsilon} ^{sy} }{10 ^{-10}} erg/cm^{2}/s\,\,\,and\,\,\, D=10D_1=[\Gamma(1-\beta Cos\theta_{obs})]^{-1}
\end{displaymath}
synchrotron peak frequency and  size of emission region
\begin{displaymath}
 \epsilon _{sy,-6} = \frac{ \epsilon  _{sy}} {10 ^{-6}}, \epsilon _{sy}= \frac{h \nu _{sy}}{m_{e} c^{2}}, R_{15} = \frac{R_{B}}{10 ^{15}} cm
\end{displaymath}
and the luminosity distance, 
\begin{displaymath}
d_{27} = \frac{d_{L}}{10^{27}} cm
\end{displaymath}
 
The size of the emission region can be estimated by taking the shortest time scale of variability applying the causality arguments. For the shortest timescale we have used $\Delta t \sim 35$ minutes, adopted from \citet{Chandra2012} which is the most probable shortest timescale of variations in the polarized flux observed during 2011, February 02-03. In our case,  size of the emission region $R_{B}$ is $(1.05) \times 10^{15} $cm, taking a typical value of Doppler factor ($\delta$) as 20. The  luminosity distance is estimated by using  cosmology calculators \citep{WrightE2006} available online.  For a spatially-flat $\Lambda$CDM cosmology, with most recent cosmological constants given by \citet{Ade2013} ($\Lambda_{M} =0.31 \pm 0.017$ ,   $\Lambda _{\nu}=1-\Lambda_{M}$  and $H_{0}=67.3 \pm 1.2$ $km\  s^{-1} Mpc^{-1}$)  the value of luminosity distance $d_{L}$  is 1022.4 Mpc or 3.1552 $\times 10 ^{27}$ cm.
  
    The estimated co-moving magnetic field using above prescription turns to be  11.86 $D_{1}^{-1} e_{B} ^{2/7} $. Here, we have obtained $p$ (=3.29) from $\alpha$ (= (p-1)/2) as estimated by fitting a power law to the synchrotron part of the SED. Assuming the equipartition of energy and Doppler boosting factor as 20, the co-moving magnetic field comes out to be $\approx $5.93 Gauss which  is typical value of the magnetic fields in BL Lac objects ($\sim$ few Gauss).

 Other quantity which we have  estimated from present study is  the mass of the black hole (BH) which 
is one of the most important parameter  in AGNs.  It controls  the accretion rate and most of the
features observed in the SEDs of AGNs. Two types of methods are used for estimating mass of the black hole: primary and secondary. The primary black hole mass estimation methods include stellar and gas kinematics, reverberation mapping and
mega-maser kinematics \citep[e.g.][]{Vestergaard2004}. The kinematics methods require high spatial resolution spectroscopy of
the host galaxy, the reverberation mapping method requires detection of the broad emission lines from BLR while mega-masers are only detectable in edge-on sources. Since blazars, in particular BL Lacs, have almost featureless  continuum and  are nearly face-on sources, all the methods mentioned above are not suitable for estimation of their black hole mass in principal. The so-called 
secondary black hole mass estimation methods are also either approximations to the reverberation mapping approach that still rely on the presence of an emission line or employ well-known empirical relations between the black hole mass and the velocity
dispersion or mass of the host galaxy's bulge.  One method which can also be used to provide a crude estimate of the mass of black hole employs time scale of the flux variability in blazars \citep{Abramowicz1982, Wiita1985, Dai2007}. The observed timescale of variability ($\Delta t$)  provides an upper limit to the mass of black hole with the assumption that the variation arises due to the processes occurring close to the black hole. The causality condition limits the size of the emission region to (R $\textless$ $\Delta t \delta$\ c/(1+z)). Combining this result with the expectation that the minimum size for such an emitting region is fairly closely related to the gravitational radius of the BH, R $\textgreater$ Rg = $GM/c^2$ \citep{Wiita1985} black hole mass can be estimated. We consider that fast variability timescale  corresponds to the perturbations in the jet plasma,  at a distance of R=5Rs, with $Rs=2GM/c^2$ as Schwartzchild radius. Then mass of mass of the BH is \citep{Dai2007},\\
\begin{equation}
 M_{BH} = \frac {c^3 \Delta t\delta}{10G(1+z)}
\end{equation}

The above expression can be used for a crude estimation of the black-hole mass of the blazars using variability timescale as a parameter in the absence of any direct method. Using  the time scale of variability considered here, central black hole mass for this source comes out to be $\approx 2.4 \times 10^{8} M_{\odot}$ which  is in  agreement with the typical black hole masses of the BL Lac objects \citep{CelottiGG08}.

  \section{Conclusions}
In this communication, we have used multi-wavelength observation data for the blazar CGRaBSJ021+1051 to understand its behaviour when it was undergoing a strong flare. The data on this source are very scanty and therefore it is not straight forward to discuss its detailed behaviour. We have used  all available data suitable for this study  in various energy regimes and found that the source shows interesting spectral behaviour. Though it is a low energy peaked BL Lac, its SED shows an equally strong IC component.

Based on the multi-wavelength light curves, we notice presence of another flare peaking sometime on 2011 January 29, with almost similar {\it Fermi}/LAT photon flux as reported for the 2011 January 23 flare. It appears to be a double flare when the decaying part of the first flare overlaps the rising part of the second one. The nature of the multi-wavelength light curves presented here show that they vary in unison with time, all fluxes peaking some time on January 29.   Such  quasi-simultaneous trend suggests that  the emissions in all the energy bands are being generated in the same part of the jet, though emitting sizes might differ with frequencies.  This behaviour also confirms the identification of the Fermi source 2FGL J0211.2+1050 as BL Lac counterpart CGRaBS J0211+1051.  That the non-thermal  jet emission dominates  and the magnetic field in the emission  region is well ordered, are    confirmed by  the  high degree of polarization observed.   It is the first detailed multi-wavelength study, including the information on the polarization on this source.  Another important conclusion drawn from this work is the confirmation of this source to be an LBL, as suggested by an earlier study based on the polarization. The synchrotron peak falls in the IR region ($\nu_{sy} \sim 1.35\times10^{14} Hz$) as shown in the SED constructed.  Synchrotron self Compton (SSC) processes can be used to explain SED and various properties of the jet emission in this source. Several parameters are estimated using canonical jet model, including  co-moving magnetic field ($\sim 5.93 $ Gauss) in the region. We estimate the mass of black hole to be $\approx 2.4 \times10^{8} M_{\odot}$.

\acknowledgments
 We thank the anonyous referee for the constructive remarks which helped to improve the quality of this communication.  This work is supported by the Department of Space, Government of India.
This research has made use of the data and/or software provided by the High Energy Astrophysics Science Archive Research Center (HEASARC), which is a service of the Astrophysics Science Division at NASA/GSFC and the High Energy Astrophysics Division of the Smithsonian Astrophysical Observatory. 
 Present work  has made use of the data from the MOJAVE database  maintained by the MOJAVE team (Lister et al., 2009, AJ, 137, 3718). This research has also made use of the NASA/IPAC Extragalactic Database (NED), operated by the Jet Propulsion Laboratory, California Institute of Technology, under contract with the National Aeronautics and Space Administration.  
 
{\it Facilities:} \facility{Swift}, \facility{Fermi (LAT)}, \facility{MIRO (ATVS, PRLPOL \& CCD)}.

 \begin{figure}
 \epsscale{.80}
 \plotone{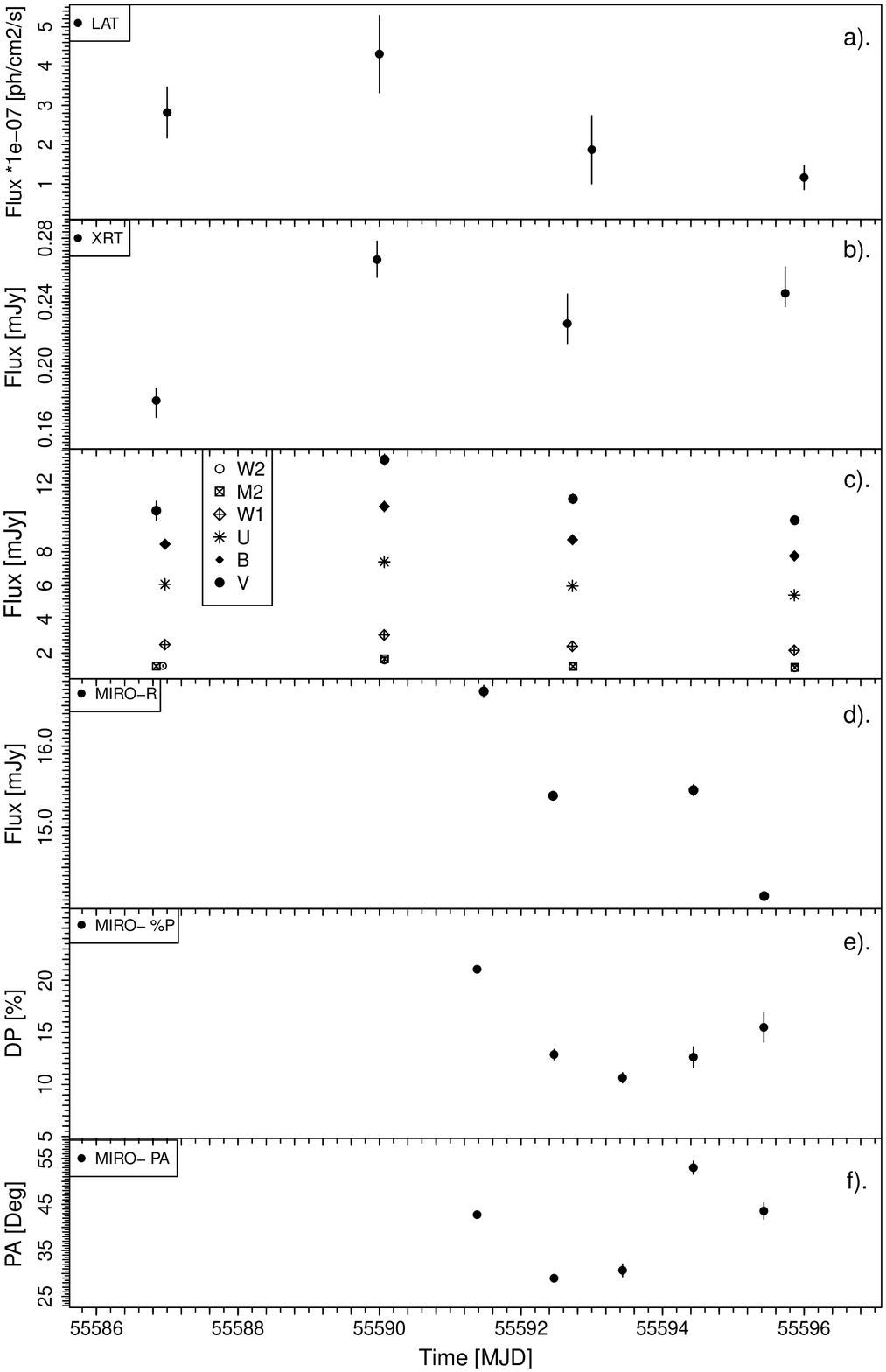}
 \caption{Multi-wavelength light curve for blazar CGRaBSJ0211+1051. (a) $\gamma$ -ray flux from {\it Fermi}/LAT, averaged over 3-days, (b) Swift-XRT x-ray integrated flux between 2 to 10 keV, (c)  Multi-bands fluxes as measured by UVOT on-board Swift, (d) daily averaged R-band flux from the ATVS-MIRO measurements, (e) \&(f) Degree of polarization \& position angle during 2011 Jan 30 to Feb 3 \citep{Chandra2012}.
 \label{fig1}}
 \end{figure}
 
 \clearpage

\begin{figure}
\epsscale{0.90}
\plotone{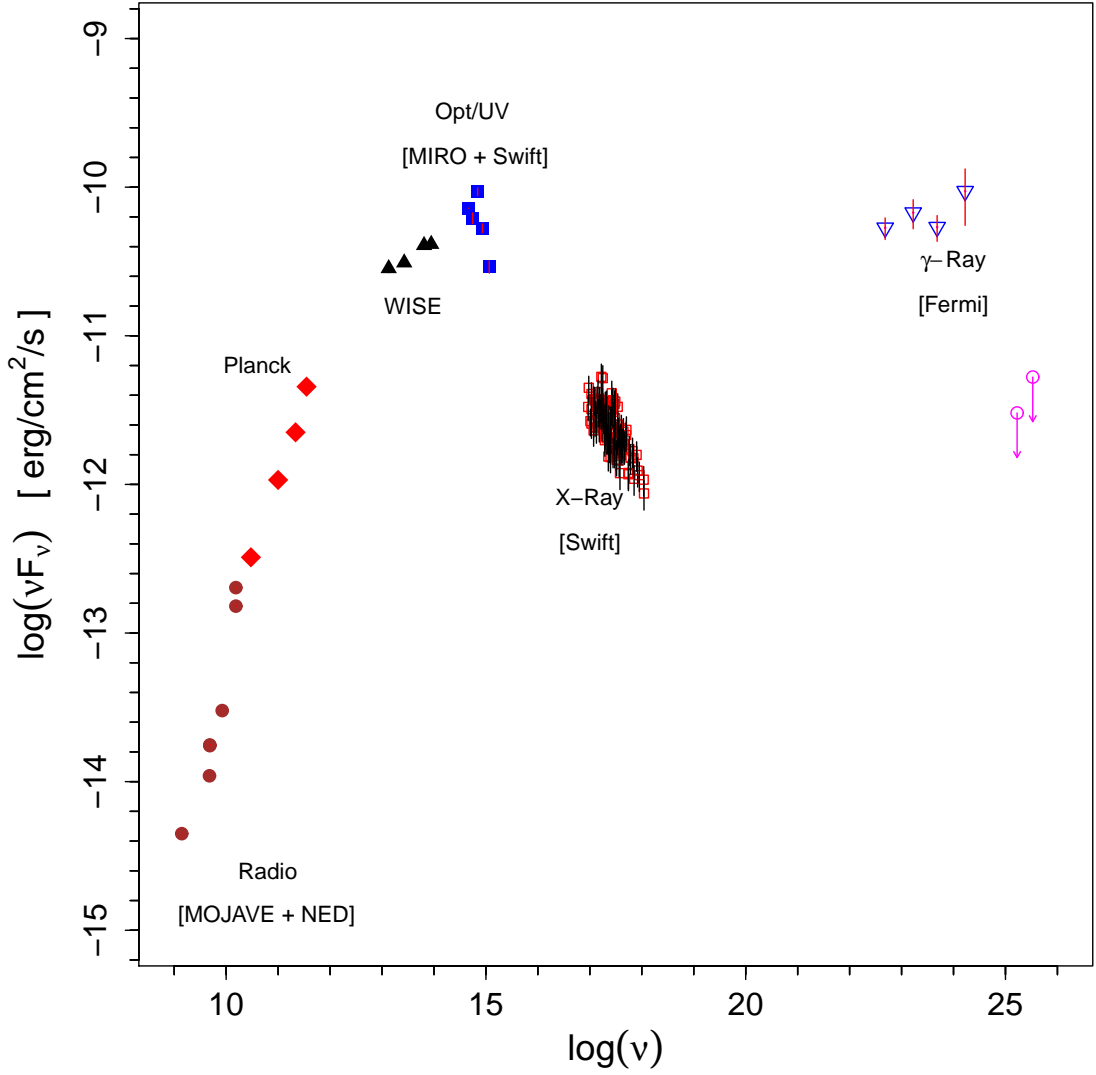}
\caption{Broadband Spectral Energy Distribution  for CGRaBSJ0211+1051 during flaring phase. The legends and annotations describe  the sources of data used in generating SED.  All fluxes are corrected for Galactic extinction. 
\label{fig2}}
\end{figure}

%%=========================================================================
%-----------------------------Swift Observations--------------
%=========================================================================

\clearpage
\begin{table}
\tabletypesize{0.2}
\rotate
\caption{Swift Observations for CGRaBS J0211+1051}
\tablewidth{0pt}
\label{tbl-1}
\begin{tabular}{|l|ll|ll|}
\hline
 Obsid&\multicolumn{2}{c|}{XRT}&\multicolumn{2}{c|}{UVOT}\\
%\cline{1-7}
&Date \& Time &Exp(s)& Time &Exp(s)\\
\hline
$00039111003$ & 2011-01-25 20:16:26 & 3919.53 & 20:17:02 & 16.52 \\
$00039111004$ & 2011-01-28 23:10:23 & 3864.32 & 23:12:41 & 65.72  \\
$00039111005$ & 2011-01-31 15:41:59 & 3449.26 & 15:45:42 & 107.05 \\
$00039111006$ & 2011-02-03 17:34:17 & 3932.49 & 17:36:27 & 60.80 \\
\hline
\end{tabular}

\end{table}
%\clearpage

%=========================================================================
%-----------------------------Swift XRT-PCA Spectrum Related--------------
%=========================================================================

\begin{deluxetable}{|ccrrrrrl|}
\tabletypesize{\scriptsize}
\rotate
\tablecaption{The spectral index ($\alpha$), associated uncertainity ($\sigma_{\alpha}$) and other parameters obtained from the  XRT spectrum fitting with  nH = 6.75 $\times$ $10^{20} cm^{-2}$.}
\tablewidth{0pt}
\label{tbl-3} 
\tablehead{ \colhead{ObsID} &   \colhead{START TIME (UT)} & \colhead{$\alpha$} & \colhead{$\sigma_{\alpha}$} & \colhead{norm} & \colhead{Unc. norm} & \colhead{${\chi}^2$/${\nu}$} & \colhead{$\nu$} }
%&&$\alpha$&$\sigma_{\alpha}$&N&$\sigma_{N}$&& \\
 
\startdata

00039111003 & 2011-01-25 20:16:26  & 2.4  & 0.09 &  1.1E-03  & 1.0E-04 & 1.02 & 17 \\
00039111004 & 2011-01-28 23:10:23  & 2.4  & 0.07 &  1.7E-03  & 1.2E-04 & 0.99 & 25 \\
00039111005 & 2011-01-31 15:41:59  & 2.3  & 0.09 &  1.4E-03  & 1.2E-04 & 1.27 & 18  \\
00039111006 & 2011-02-03 17:34:17  & 2.4  & 0.07 &  1.5E-03  & 1.1E-04 & 0.93 & 25 \\

\enddata
\end{deluxetable}

%\clearpage

%=========================================================================
%-----------------------------R-Band Optical Photometry-------------------
%=========================================================================

\begin{deluxetable}{|rcrrrrrrrrl|}
\tabletypesize{\scriptsize}
\rotate
\tablecaption{Galactic extinction corrected R-band  magnitues for CGRABsJ0211+1051 and comparison stars used.}
\tablewidth{0pt}
\label{tbl-4}
\tablehead{ \colhead{Date} &   \colhead{Time (MJD)} & \colhead{S (mag)} &
\colhead{$e_{s}$} & \colhead{$\sigma_{S}$} & \colhead{ C1 (mag)} &
\colhead{$\sigma_{C1}$} & \colhead{C2 (mag)} & \colhead{$\sigma_{C2}$} &
\colhead{C3 (mag)} & \colhead{$\sigma_{C3}$}
}

\startdata

2011 Jan 29 & 55591.47 & 13.23 & 0.005 & 0.039 & 13.54 & 0.037 & 12.10 & 0.033 & 12.29 & 0.039 \\
     Jan 31 & 55592.45 & 13.32 & 0.004 & 0.030 & 13.54 & 0.051 & 12.10 & 0.043 & 12.28 & 0.046 \\
     Feb 02 & 55594.44 & 13.31 & 0.005 & 0.063 & 13.53 & 0.059 & 12.08 & 0.059 & 12.27 & 0.063 \\
     Feb 03 & 55595.44 & 13.41 & 0.003 & 0.036 & 13.53 & 0.041 & 12.09 & 0.037 & 12.28 & 0.038  \\

\enddata
\end{deluxetable}


\clearpage

\begin{thebibliography}{}
\bibitem[{Abdo} {\em et~al.}(2009)]{Abdo2009}
{Abdo}, A.~A., {Ackermann}, M., {Ajello}, M., et~al. 2009, {\em \apj\/},{\bf 699}, 976--984.
\bibitem[{Abdo} {\em et~al.}(2010)]{Abdo2010}
{Abdo}, A.~A., {Ackermann}, M., {Ajello}, M., et~al. 2010, {\em \nat\/},{\bf 463}, 919--923.
\bibitem[{Abramowicz} and {Nobili}(1982)]{Abramowicz1982}
{Abramowicz}, M.~A. and {Nobili}, L. (1982).{\em \nat\/}, {\bf 300}, 506.
\bibitem[{Ade} {\em et~al.}(2011)]{Planck11}
{Ade}, P.~A.~R., {Aghanim}, N.,  {Arnaud}, M., and {Planck Collaboration}, et~al., 2011, {\em \aap\/},{\bf 536, A2}.
\bibitem[Ade  et al.(2013)]{Ade2013} 
 Ade, P.~A.~R., Aghanim, N., N.  and {Planck Collaboration}, et al.\ 2013, arXiv:1303.5076
\bibitem[{Agudo} {\em et~al.}(2011)]{Agudo2011}
{Agudo}, I. and {Marscher}, A.~P. and {Jorstad}, S.~G., {\em et~al.}, 2011, {\em \apjl\/}, {\bf 735}, L10.
\bibitem[{Andruchow} {\em et~al.}(2005){Andruchow}, {Romero}, and {Cellone}]{Andruchow2005}
{Andruchow}, I., {Romero}, G.~E., and {Cellone}, S.~A. (2005).{\em \aap\/}, {\bf 442}, 97--107.
\bibitem[{Atwood} {\em et~al.}(2009)]{Atwood2009}
{Atwood}, W.~B. and {Abdo}, A.~A. and {Ackermann}, M., et~al. 2009, {\em \apj\/}, {\bf 697}, 1071--1102.
\bibitem[{Bajaja} {\em et~al.}(2005)]{Bajaja05}
{Bajaja}, E., {Arnal}, E. M., {Larrarte}, J. J., et~al. 2005, {\em \aap\/}, {\bf 440}, 767--773.
\bibitem[{Barthelmy} {\em et~al.}(2005)]{Barthelmy2005}
{Barthelmy}, S.~D. and {Barbier}, L.~M. and {Cummings}, J.~R., et~al. 2005, {\em \ssr\/}, {\bf 120}, 143--164.
\bibitem[{Bessel}{\em et~al.}(1979)]{bessel1979}
{Bessel}, M~.S. , 1979, {\em Publications of the Astronomical society of the pacific}, {\em 91}, 589-607.
\bibitem[{B{\l}a{\.z}ejowski} {\em et~al.}(2000){B{\l}a{\.z}ejowski}, and {Sikora},{Moderski}]{Blazejowski2000}
{B{\l}a{\.z}ejowski}, M., {Sikora}, M., {Moderski}, R., and {Madejski}, G.~M. 2000, {\em \apj\/}, {\bf 545}, 107--116.
\bibitem[{Bloom} and {Marscher}(1996){Bloom} and {Marscher}]{BloomMarscher1996}
{Bloom}, S.~D. and {Marscher}, A.~P. 1996, {\em \apj\/}, {\bf 461}, 657.
\bibitem[{Bloom} (2008)]{Bloom2008} 
{Bloom}, S.~D., 2008, {\em \aj\/}, {\bf 136}, 1533. 
\bibitem[{B{\"o}ttcher}(2007){B{\"o}ttcher}]{bottcher2007}
{B{\"o}ttcher}, M. 2007, {\em \apss\/}, {\bf 309}, 95--104.
\bibitem[{Burrows} {\em et~al.}(2005)]{Burrows2005}
{Burrows}, D.~N. and {Hill}, J.~E. and {Nousek}, J.~A., {\em et~al.,} 2005, {\em \ssr\/}, {\bf 120}, 165--195.
\bibitem[{Cardelli} {\em et~al.}(1989){Cardelli}, {Clayton}, and {Mathis}]{Cardelli1989}
{Cardelli}, J.~A., {Clayton}, G.~C., and {Mathis}, J.~S. 1989, {\em \apj\/}, {\bf 345}, 245--256.
\bibitem[{Chandra}  et al.(2011a){Chandra}, and {Baliyan} and {Ganesh}]{Chandra2011}
{Chandra}, S., {Baliyan}, K.~S.,  {Ganesh}, S.,  \& {Joshi}, U.~C. 2011, {\apj\/}, 731, 118.
\bibitem[{Chandra} {\em et~al.}(2012){Chandra}, and {Baliyan}, and {Ganesh}]{Chandra2012}
{Chandra}, S., {Baliyan}, K.~S., {Ganesh}, S., and {Joshi}, U.~C. 2012, {\em \apj\/}, {\bf 746}, 92.
\bibitem[{Carrasco} et al. (2013){Carrasco}, and {Porras}, and {Recillas}]{Carrasco2013}
{Carrasco}, L., {Porras}, A., {Recillas}, E. et~al., 2013,  {\em The Astronomer's Telegram\/}, {\bf 5570}, 1.
\bibitem[{D'Ammando}(2011){D'Ammando}]{AmmandoAtel2011}
{D'Ammando}, F. 2011, {\em The Astronomer's Telegram\/}, {\bf 3120}, 1.
\bibitem[D'Ammando et al.(2011)]{AmmandoAtel2011a} 
{D'Ammando}, F., {Sokolovsky}, K.~V., \& {Hoversten}, E.\ 2011,{\em The Astronomer's Telegram\/}, {\bf 3129}, 1.
\bibitem[{Dai} et~al.(2007)]{Dai2007}
{Dai}, H., {Xie}, G.~H., {Zhou}, S.~B., et~al, 2007, {\em \aj\/}, {\bf 133}, {\bf 2187}
\bibitem[{Dermer} {\em et~al.}(1992){Dermer}, {Schlickeiser}, and {Mastichiadis}]{Dermer1992}
{Dermer}, C.~D., {Schlickeiser}, R., and {Mastichiadis}, A. 1992, {\em \aap\/}, {\bf 256}, L27--L30.
\bibitem[{Djorgovski} et al.(2011)]{Djorgovski2011}
{Djorgovski}, S.~G. and {Drake}, A.~J. and {Mahabal}, A.~A., et~al. 2011,   Atel  3133, 1.
\bibitem[{Fan} {\em et~al.}(1997){Fan}, {Okudaira}, {Lin}, and {Xie}]{Fan1997}
{Fan}, J.~H., {Okudaira}, A., {Lin}, R.~G., and {Xie}, G.~Z. 1997, {\em \apss\/}, {\bf 253}, 275--284.
\bibitem[{Foschini} {\em et~al.}(2010)]{FoschiniL10}
{Foschini}, L. and {Tagliaferri}, G. and {Ghisellini}, G., et~al. 2010, {\em \mnras\/}, {\bf 408}, 448--451.
\bibitem[{Foschini} {\em et~al.}(2011){Foschini}, {Ghisellini}, and {Tavecchio}]{FoschiniL11}
{Foschini}, L., {Ghisellini}, G., {Tavecchio}, F., {Bonnoli}, G., and {Stamerra}, A. 2011, {\em \aap\/}, {\bf 530}, A77.
\bibitem[{Fossati} {\em et~al.}(1998){Fossati}, {Maraschi}, and {Celotti}]{Fossati1998}
{Fossati}, G., {Maraschi}, L., {Celotti}, A., {Comastri}, A., and {Ghisellini}, G. 1998, {\em \mnras\/}, {\bf 299}, 433--448.
\bibitem[{Celotti}(2008){Celotti} and {Ghisellini}]{CelottiGG08}
{Celotti}, A., {Ghisellini}, G., 2008, \mnras, 385, 283.
\bibitem[Ghisellini et~al. (1985)]{GG85} 
{Ghisellini}, G. and {Maraschi}, L. and {Treves}, A., et~al., 1985, \aap, 146, 204.
\bibitem[{Gorbovskoy}  et al.(2011)]{Gorbovskoy11} 
{Gorbovskoy}, E. and {Balanutsa}, P. and {Lipunov}, V., et al. 2011, Atel 3134, 1.
\bibitem[{Greiner} {\em at~al.}(2008)]{Griener2008}
{Greiner}, J. and {Bornemann}, W. and {Clemens}, C., et al. 2008, {\em \pasp\/}, {\bf 120, 405..424}.
\bibitem[Gregory \& Condon(1991)]{Gregory1991}
{Gregory}, P.~C., \& {Condon}, J.~J.\ 1991, {\em \apjs\/},{\bf 75}, 1011. 
\bibitem[Griffith et al.(1991)]{Griffith1991} 
{Griffith}, M., {Heflin}, M., {Conner}, S., {Burke}, B., \& {Langston}, G.\ 1991,{\em \apjs\/}, 75, 801. 
\bibitem[{Grigoreva} et al. (2013){Grigoreva}, and {Krushinsky}, and {Pruzhinskaya}]{Grigoreva2013}
{Grigoreva}, E., {Krushinsky}, V., {Pruzhinsoya}, M. et~al., 2013,  {\em The Astronomer's Telegram\/}, {\bf 5588}, 1.
\bibitem[{Hartman} et~al. (1999)]{Hartman1999}
{Hartman}, R.~C., {Bertsch}, D.~L., {Bloom}, S.~D., et al., 1999, {\em \apjs\/}, {\bf 123}, 79.
\bibitem[{Healey} et al.(2007)]{Healey2007}
{Healey}, S.~E. and {Romani}, R.~W. and {Taylor}, G.~B., et al. 2007, {\apjs\/}, 171, 61.
\bibitem[{Healey} et al.(2008) ]{Healey2008}
{Healey}, S.~E. and {Romani}, R.~W. and {Cotter}, G., et al.  2008, {\apjs\/},  175, 97.
\bibitem[{Heidt} and {Nilsson}(2011){Heidt} and {Nilsson}]{HeidtNilsson2011}
{Heidt}, J. and {Nilsson}, K. 2011, {\em \aap\/}, {\bf 529}, A162.
\bibitem[{Hill} {\em et~al.}(2004){Hill}, {Gulliver}, and {Adelman}]{Hill2004}
{Hill}, G., {Gulliver}, A.~F., and {Adelman}, S.~J. 2004, {\em The A-Star Puzzle\/}, volume 224 of {\em IAU Symposium\/}, pages 35--42.
\bibitem[{Ikejiri} {\em et~al.}(2009)]{Ikejiri2009}
{Ikejiri}, Y. and {Uemura}, M. and {Sasada}, M., et~al. 2009, {\em ArXiv e-prints\/}.
\bibitem[{Jannuzi} {\em et~al.}(1993){Jannuzi}, {Green}, and {French}]{Jannuzi1993}
{Jannuzi}, B.~T. and {Green}, R.~F. and {French}, H., 1993, {\em \apj\/}, {\bf 404}, 100--111.
\bibitem[{Kalberla} {\em et~al.}(2005)]{Kalberla05}
{Kalberla}, P.~M.~W. and {Burton}, W.~B. and {Hartmann}, D., et~al. 2005, {\em \aap\/}, {\bf 440}, 775--782.
\bibitem[{Lawrence} et al.(1986)]{Lawrence1986}
{Lawrence}, C.~R. and {Bennett}, C.~L. and {Hewitt}, J.~N., et al.  1986 {\apjs\/}, 61, 105.
\bibitem[{Lister} {\em et~al.}(2009)]{ListerMOJAVE2009}
{Lister}, M.~L. and {Aller}, H.~D. and {Aller}, M.~F., et~al, 2009, {\em \aj\/}, {\bf 137}, 3718--3729.
\bibitem[{Marscher} et. al. (2010)]{Marscher2010}
{Marscher}, A.~P. and {Jorstad}, S.~G. and {Larionov}, V.~M., et. al. 2010, {\em \apj\/}, {\bf 710}, 126.
\bibitem[{Meisner} and {Romani}(2010){Meisner} and {Romani}]{MeisnerRomani2010}
{Meisner}, A.~M. and {Romani}, R.~W. 2010, {\em \apj\/}, {\bf 712}, 14--25.
\bibitem[{Nesci}(2011){Nesci}]{NesciAtel2011}
{Nesci}, R. 2011, {\em The Astronomer's Telegram\/}, {\bf 3127}, 1.
\bibitem[{Nolan} {\em et~al.}(2012)]{nolan2012}
{Nolan}, P.~L., {Abdo}, A.A., {Ackermann}, M., et~al. 2012, {\em \apjs\/},{\bf 199}, 31.
\bibitem[{Orienti}, et. al. (2013)]{Orienti2013}
{Orienti}, M. and {Koyama}, S. and {D'Ammando}, F., et~al., 2013, {\em \mnras\/}, {\bf 428}, 2418.
\bibitem[{Poole} {\em et~al.}(2008)]{Poole2008}
{Poole}, T.~S. and {Breeveld}, A.~A. and {Page}, M.~J., et~al. 2008, {\em \mnras\/}, {\bf 383}, 627--645.
\bibitem[{Rau}{\em et~al.}(2012)]{Rau2012}
{Rau}, A. and {Schady}, P. and {Greiner}, J., et~al. 2012, {\em \aap\/},{\bf 538, A26}.
\bibitem[{Roming} {\em et~al.}(2005)]{Roming2005}
{Roming}, P.~W.~A. and {Kennedy}, T.~E. and {Mason}, K.~O., et~al. 2005, {\em \ssr\/}, {\bf 120}, 95--142.
\bibitem[{Sambruna} {\em et~al.}(2009){Sambruna}, {Donato}, and {Ajello}]{Sambruna2009}
{Sambruna}, R.~M., {Donato}, D., {Ajello}, M., {Maraschi}, L., and {the GSFC BAT Team} 2009, {\em ArXiv e-prints\/}.
\bibitem[{Serkowski}(1974){Serkowski}]{Serkowaski1974}
{Serkowski}, K. 1974, {\em IAU Colloq. 23: Planets, Stars, and Nebulae: Studied with Photopolarimetry\/}, page 135.
\bibitem[{Sikora} {\em et~al.}(1994){Sikora}, {Begelman}, and {Rees}]{Sikora1994}
{Sikora}, M., {Begelman}, M.~C., and {Rees}, M.~J. 1994, {\em \apj\/}, {\bf 421}, 153--162.
\bibitem[{Sikora} {\em et~al.}(2009){Sikora}, {Stawarz}, and {Moderski}]{sikora2009}
{Sikora}, M., {Stawarz}, {\L}., {Moderski}, R., {Nalewajko}, K., and {Madejski}, G.~M. 2009, {\em \apj\/}, {\bf 704}, 38--50.
\bibitem[{Sokolov} {\em et~al.}(2004){Sokolov}, {Marscher}, and {McHardy}]{SokolovMarscher2004}
{Sokolov}, A., {Marscher}, A.~P., and {McHardy}, I.~M. 2004, {\em \apj\/}, {\bf 613}, 725--746.
\bibitem[{Snellen} et al.(2002)]{Snellen2002}
{Snellen}, I.A.G., {McMahon}, R.G., {Hook}, I.M.F., \& {Browne}, I.W.A. 2002, {\mnras\/}, 329, 700.
\bibitem[{Tommasi} {\em et~al.}(2001)]{Tommasi2001}
{Tommasi}, L. and {Palazzi}, E. and {Pian}, E., et~al. 2001, {\em \aap\/}, {\bf 376}, 51--58.
\bibitem[{Urry} and {Mushotzky}(1982){Urry} and {Mushotzky}]{UrryMushotzsky1982}
{Urry}, C.~M. and {Mushotzky}, R.~F. 1982, {\em \apj\/}, {\bf 253}, 38--46.
\bibitem[{Urry} and {Padovani}(1995){Urry} and {Padovani}]{UrryPadovani1995}
{Urry}, C.~M. and {Padovani}, P. 1995, {\em \pasp\/}, {\bf 107}, 803.
\bibitem[{Urry} and {Padovani}(2000){Urry} and {Padovani}]{UrryPadovani2000}
{Urry}, M. and {Padovani}, P. 2000, {\em \pasp\/}, {\bf 112}, 1516--1518.
\bibitem[{Vestergaard} (2004) {Vestergaard}, M.,]{Vestergaard2004}
{Vestergaard}, M., 2004, {\em Astronomical Society of the Pacific Conference Series}, {\bf 311}, {\bf 69}
\bibitem[{Wiita} (1985){Wiita}, P.~J.]{Wiita1985}
{Wiita}, P.~J., 1985, {\em \physrep\/}, {\bf 123}, {\bf 117-213}
\bibitem[{Wright}(2006) {Wrigth}, E,~L.,]{WrightE2006}
{Wrigth},E,~L. 2006, {\em \pasp\/}, {\bf 118}, 1711-1715. 
\bibitem[{Wright} (2010) {Wrigth}, E,~L.]{Wright2010}
{Wrigth},E,~L. 2010, {\em \aj\/}, {\bf 140}, 1868-1881. 

\end{thebibliography}
\end{document}